\documentclass[10pt]{article}
\usepackage{amsmath}
\usepackage{amssymb}
\usepackage{graphicx}

\newcommand{\ve}[1]{\mbox{\boldmath$#1$}}

\begin{document}

\setcounter{figure}{0}
\setcounter{table}{0}
\setcounter{footnote}{0}
\setcounter{equation}{0}

\vspace*{0.5cm}

\noindent {\Large Towards sub-microarcsecond models for relativistic astrometry}
\vspace*{0.7cm}

\noindent\hspace*{1.5cm} S.\ ZSCHOCKE, S.A.\ KLIONER, M.H.\ SOFFEL \\

\noindent\hspace*{1.5cm} Lohrmann Observatory\\
\noindent\hspace*{1.5cm} Helmholtzstrasse 10, D-01062 Dresden, Germany\\
\noindent\hspace*{1.5cm} e-mail: michael.soffel@tu-dresden.de\\

\vspace*{0.5cm}

\noindent {\large ABSTRACT.} Astrometric space missions like Gaia have
stimulated a rapid advance in the field of relativistic astrometry. Present
investigations in that field aim at accuracies significantly less than a
microarcsecond. We review the present status of relativistic astrometry. As
far as the problem of light propagation is concerned we face two problems:
the form of the BCRS metric and solutions to the light-ray equation. Finally, 
work in progress in that field is briefly mentioned.

\vspace*{1cm}

\noindent {\large 1. INTRODUCTION}

\smallskip

The Hipparcos astrometric space mission has determined positions (proper
motions) of some 120000 stars with a precision of a milliarcsecond (mas/y).
The forthcoming mission Gaia is expected to reach a level of up to some
$\mu$as for one billion stars depending on stellar brightness. Proposed 
missions like the 'Nearby Earth Astrometric Telescope' (NEAT) envisage
an accuracy of $50$ nanoarcseconds (nas).  This stunning progress in astrometry
implies the necessity to formulate appropriate relativistic astrometric
models with an intrinsic accuracy of $1\,{\rm nas}$. One is still far from that
goal but there has been a lot of work in that direction.

Any relativistic astrometric model based on Einstein's theory of gravity
employs one or several different reference systems (4-dimensional coordinate
systems) to describe the location and motion of gravitating bodies and the
light trajectory from the emitter to the observer. In one of these well
related coordinate systems one has to formulate the astrometric observable as
a coordinate independent quantity (i.e., as a scalar). A model for a concrete
astrometric mission will contain a certain set of coordinate-dependent
parameters that have to be fitted from observational data. The reference
system then becomes the corresponding reference frame, materialized e.g., by
a stellar (or quasar) catalog.

If physically relevant local coordinates, co-moving with the observer are
introduced, then it might be possible to derive observables from coordinate
quantities as it is the case in the Gaia Relativistic Model (GREM) developed by Klioner (2003a).
An astrometric model thus involves the following constructions: i) one or
several space-time reference systems, i.e., space-time coordinates and the
corresponding metric tensor, ii) the trajectories of light-rays, iii) the
trajectories of the observer and gravitating bodies,  and iv) the calculation
of astrometric observables. This contribution focuses on the first two
aspects i) and ii).

\vspace*{0.7cm}

\noindent {\large 2. APPROXIMATION METHODS FOR REFERENCE SYSTEMS}

\smallskip

To construct a space-time reference system with a metric tensor as solution
of Einstein's field equations  for real high precision astrometric
observations one resorts to approximation schemes, either to a post-Newtonian
hierarchy (weak field, slow motion) or to the post-Minkowskian approximation
(weak field). For light rays the post-Newtonian (PN) metric is of the form
\begin{eqnarray}
g_{00} &=& -1 + \frac{2 w}{c^2} \, , \qquad g_{0i} = 0 \, , \qquad g_{ij} =
\left( 1 + \frac{2 w}{c^2} \right) \delta_{ij}\,,
\end{eqnarray}

\noindent
where $w$ is the gravitational potential. The corresponding post-post
Newtonian (2PN) metric for light rays can be written in the form
\begin{eqnarray}
g_{00} &=& -1 + \frac{2w}{c^2} - \frac{2 w^2}{c^4} \, , \quad g_{0i} = -
\frac{4 w^i}{c^3} \,, \quad g_{ij} =  \left( 1 + \frac{2 w}{c^2} + \frac{2
w^2}{c^4}\right) \delta_{ij} + \frac{4}{c^4} q_{ij} \, , \label{met}
\end{eqnarray}

\noindent
where $w^i$ is the gravito-magnetic potential induced by moving or rotating
masses.
For one body at rest (rotating, vibrating, with arbitrary shape and
decomposition) the exterior metric is known for both
the post-Newtonian (Blanchet \& Damour, 1989) and the post-Minkowskian case
(Damour \& Iyer, 1991), and it has been demonstrated that the metric in both cases is
determined by only two families of multipole moments, $M_L$ (mass-moments) and $S_L$ (spin-moments).

In what follows we will give an overview of the present status of the theory of light propagation. 

\vspace*{0.7cm}

\noindent {\large 3. LIGHT PROPAGATION IN THE FIELD OF MASS MONOPOLES}

\smallskip

{\bf Light propagation in the field of mass monopoles with constant
velocity:}
Explicit post-Newtonian solutions for the light propagation in the case of
uniformly moving bodies, where the position of the body is given by 
$\ve{x}_A(t)=\ve{x}_A^{\rm eph}(t_A^0)+\dot{\ve x}_A^{\rm eph}(t_A^0)\left(t-t_A^0\right)$, 
were derived by Klioner (1989); here $\ve{x}_A^{\rm eph}$ and $\dot{\ve x}_A^{\rm eph}$ are the actual 
position and velocity of body $A$ taken from an ephemeris for some instant of time $t_A^0$. 
Following a suggestion by
Hellings (1986), Klioner \& Kopeikin (1992) have argued that in order to
minimize the errors in the light propagation the free parameter $t_A^0$ 
should be chosen to coincide with the moment of closest   
approach between the body and the light ray. Furthermore, Klioner (2003b) has
suggested a straightforward way to calculate the effect of uniform
translational motion of a  body on the light propagation by a Lorentz
transformation of the light trajectory in a reference system where the body
is at rest. In this way Klioner (2003b) has derived a post-Minkowskian
solution for the light propagation in the field of a mass monopole moving
with constant velocity. It has been demonstrated that the more general
solution of Kopeikin \& Sch\"afer (1999) can be reproduced in the limiting
case of uniform motion.

 {\bf Light propagation in the field of arbitrarily moving mass monopoles:}
A rigorous solution of the problem in the first post-Minkowskian approximation has been found
by Kopeikin \& Sch\"afer (1999), where the geodetic equations for photons are integrated using retarded 
potentials.
The numerical accuracy of various approaches has been investigated by Klioner \& Peip (2003).
Especially, Klioner \& Peip (2003) have numerically compared various available
solutions for the light propagation for  observations made within
the Solar system. The authors used both artificial orbits for
deflecting bodies as well as planetary trajectories taken from JPL
solar system ephemerides. It has been demonstrated that the simple
solution obtained in (Klioner, 1991) and (Klioner \& Kopeikin, 1992) is sufficient
for an accuracy of about $2\,{\rm nas}$, provided that $\ve{x}_A^{\rm eph}$ and $\dot{\ve x}_A^{\rm eph}$ 
are taken in the optimal way.

 {\bf 2-PN Light propagation in the field of mass monopoles at
rest:}
The light trajectories in the Schwarzschild field, that means in the field of
a single mass monopole at rest, can be found in an analytically closed
form as it has been demonstrated at the first time by Hagihara (1931); 
for a re-derivation we refer to Chandrasekhar (1983). 
However, this exact analytical solution is not convenient for data reduction of astrometric observations, 
since the light curve is not given by an explicit time dependence of 
the coordinates of the photon $x(t), y(t)$ but only implicitly in terms of $y(x)$.  

From a practical point of view,  post-post-Newtonian effects in the light
propagation in the Schwarz\-schild field have been considered by many authors.   
An important progress has been made by Brumberg (1991) who has found an
explicit post-post-Newtonian solution for light trajectories in the
Schwarzschild field as function of coordinate time in a number of coordinate
gauges. Generalizations of that solution for the case of the parametrized
post-post-Newtonian metric have been given by Klioner \& Zschocke (2010). The
latter authors have investigated in great detail the numerical magnitudes of
various post-post-Newtonian terms and formulated practical algorithms for
highly-effective computation of the post-post-Newtonian effects. It has been
demonstrated that the so-called enhanced post-post-Newtonian terms are due to
a physically inadequate choice of the parametrization of the light rays; see
also Bodenner \& Will (2003).

Two alternative approaches to the calculation of propagation times and directions of light rays have been formulated
recently. Both approaches allow one to avoid explicit integration of the geodetic equations for light rays. The first
approach (Le Poncin-Lafitte, Linet \& Teyssandier, 2004; Teyssandier \& Le Poncin-Lafitte, 2008) is based on the use of Synge's world
function. Several applications of this approach have been published: higher post-Newtonian approximations in spherically
symmetric gravitational fields and post-Newtonian effects in the gravitational field with multipole moments. Another
approach based on the eikonal concept  has been developed by Ashby \& Bertotti (2010) to investigate the light
propagation in the field of a spherically symmetric body. All the results of these authors confirm the conclusions and
formulas obtained in Klioner \& Zschocke (2010).

\newpage

{\bf 2-PN Light propagation in the field of moving mass monopoles:}
There are only very limited results dealing with moving deflecting bodies in
the post-post-Newtonian approximation. 
Especially, Br\"ugmann (2005) has investigated some effects of the light propagation in
the post-post-Newtonian gravitational field of a system of two bodies, where two
important approximations are used. First, both the light source and
the observer are assumed to be located at infinity in an asymptotically flat
space. Second, some of the results were obtained in form of an expansion in
powers of the ratio between the distance between two bodies and the impact
parameter of the light ray with respect to the center of mass of the two-body
system. These assumptions imply, however, that the results are not
applicable to observations in the solar system.

\vspace*{0.7cm}

\noindent {\large 4. LIGHT PROPAGATION IN THE FIELD OF MASS QUADRUPOLES}

\smallskip

{\bf Light propagation in the quadrupole field of bodies at rest:} 
Analytical solutions of light deflection in a quadrupole gravitational field
have previously been investigated by many authors. However, for the first
time the full analytical solution for the light trajectory in a quadrupole
field has been obtained by Klioner (1991), where an explicit time dependence
of the coordinates of a photon and the solution of the boundary value problem
for the geodetic equation has been obtained. These results were confirmed by a
different approach in Le Poncin-Lafitte \& Teyssandier (2008), while a
simplified expression with $\mu{\rm as}$ accuracies has been derived in
Zschocke \& Klioner (2011).

{\bf Light propagation in the quadrupole field of arbitrarily moving bodies:}
The light-deflection at moving massive bodies, having monopole  and
quadrupole structure, has been investigated by Kopeikin \& Makarov (2007),
where the quadrupole term is taken into account in local coordinates of the
body in Newtonian approximation. Using the  harmonic gauge, the linearized
Einstein equations are inhomogeneus wave equations and a general solution is
given in terms of a multipole expansion (Thorne, 1980; Blanchet \& Damour,
1986). In Kopeikin \& Makarov (2007) the geodesic equation is rewritten into
a considerably simpler form. Using a special integration method, they
succeeded to integrate analytically the geodesic equation by neglecting all
terms that contribute by less than $1\,\mu{\rm as}$.

\vspace*{0.7cm}

\noindent {\large 5. LIGHT PROPAGATION IN THE FIELD OF BODIES WITH SPIN}

\smallskip

{\bf Light propagation in the field of bodies at rest with a spin-dipole:}
The first explicit post-Newtonian solution of the light trajectory in the
gravitational field of massive bodies at rest possessing a spin dipole has
been obtained by Klioner  (1991). This solution provides all the details of
light propagation, especially the explicit time dependence of the coordinates
of the photon and the solution of the corresponding boundary value problem.
Kopeikin (1997) has generalized the solution for the case of motionless
bodies possessing any set of time-independent spin (and mass) moments, and it
has been shown that the expression in Klioner (1991) and Kopeikin (1997)
agree with each other.

{\bf Light propagation in the field of arbitrarily moving bodies with
spin-dipole:}
Kopeikin \& Mashhoon  (2002) have derived formulas for the case of light
propagation in the field of arbitrarily moving bodies possessing mass
monopole and spin dipole.

\vspace*{0.7cm}

\noindent {\large 6. LIGHT PROPAGATION IN THE FIELD OF HIGHER MASS AND SPIN
MULTIPOLE MOMENTS}

\smallskip

{\bf Mass and spin multipole moments at rest:}
A systematic approach to the integration of light geodesic equation in the stationary post-Newtonian  
gravitational field of an isolated system of $N$ bodies having a complex but time-independent multipole  
structure has been worked out in Kopeikin (1997) and Kopeikin et al. (1999).   
Especially, the work of Kopeikin (1997) represents a generalized solution for the case of
motionless bodies possessing any set of time-independent mass and spin
moments, that is $M_L$ and $S_L$, respectively.
Later, in Kopeikin, Korobkov \& Polnarev (2006) and Kopeikin \& Korobkov (2005),
the propagation of light rays in the field of localized sources which are
completely characterized by time-dependent mass and spin multipoles, $M_L
\left(t\right)$ and $S_L \left(t\right)$, respectively, has been
investigated. 
Kopeikin, Korobkov \& Polnarev (2006) and
Kopeikin \& Korobkov (2005) have found an analytical solution for the light
propagation in such gravitating systems.

\newpage

\noindent {\large 7. WORK IN PROGRESS}

\smallskip
Presently several groups try to extend relativistic astrometry to still
higher accuracies. Our group presently concentrates on the 2PN field of
arbitrarily moving bodies endowed with arbitrary mass- and spin multipole
moments where the metric in harmonic gauge is given by (\ref{met}). There
have been first attempts to tackle this problem (e.g.,  Xu \& Wu 2003;
Minazzoli \& Chauvineau 2009) but they are far from being complete.

Problems, that have been ignored in these preliminary papers, are  
related with the internal structure of the bodies. For a single body at rest
these problems are well understood for both the post-Newtonian and the
post-Minkowskian case (Blanchet \& Damour, 1989; Damour \& Iyer, 1991) where
many structure dependent terms appear in intermediate calculations that
cancel exactly in virtue of the local equations of motion or can be
eliminated by corresponding gauge transformations. However, for the
post-linear case  the situation is still unclear. In course of our
studies for the general problem just mentioned we found that even for the
spherically symmetric case of a single body the complete derivation of the
exterior metric (the Schwarzschild metric) is interesting. In a forthcoming
paper (Klioner \& Soffel, 2014) we will show how such structure-dependent
terms cancel and one ends up with the Schwarzschild solution in harmonic
gauge.

\vspace*{0.7cm}


\noindent {\large 7. REFERENCES}

{

\leftskip=5mm
\parindent=-5mm

\smallskip

Ashby, N., Bertotti, B., 2010, "Accurate light-time correction due to a
gravitating mass", Class. Quantum Grav. 27, 145013.

Blanchet, L., Damour, T., 1986, "Radiative gravitational fields in general
relativity: I. General Structure of the field outside the source", Phil.
Trans. R. Soc. London A 320, 379.

Blanchet, L., Damour, T., 1989, "Post-Newtonian generation of gravitational
waves", Annales de I'lnstitut Henri Poincar\'e I Physique Theorique 50, 377.

Bodenner, J., Will, C.M., 2003, "Deflection of light to second order: A tool
for illustrating principles of general relativity", Am. J. Phys. 71, 770.

Br\"ugmann, M.H., 2005, "Light deflection in the post-linear gravitational
field of bounded pointlike masses", Phys. Rev. D 72, 024012.

Brumberg, V.A., 1991, "Essential Relativistic Celestial Mechanics",  Adam
Hilger, Bristol

Chandrasekhar, S., 1983, "The mathematical Theory of Black Holes", Clarendon
Press, Oxford.

Damour, T., Iyer, B.R., 1991, "Multipole analysis for electromagnetism and
linearized gravity with irreducible Cartesian tensors", Phys. Rev. D 43,
3259.

Hagihara Y., 1931, "Theory of the relativistic trajectories in a gravitational field of Schwarzschild", 
Jap. J. Astron. Geophys. 8, 67. 

Hellings, R.W., 1986, "Relativistic Effects in Astronomical Timing
Measurements", AJ 91, 650.

Klioner, S.A., 1989, "Propagation of the Light in the Barycentric Reference
System considering the Motion of the Gravitating Masses", Communications of
the Institute of Applied Astronomy No 6, 21.

Klioner, S.A., 1991, "Influence of the quadrupole field and rotation of
objects on the light propagation", Sov. Astron. 35, 523.

Klioner, S.A., 2003a, "Practical Relativistic Model of Microarcsecond
Astrometry in Space", AJ 125, 1580.

Klioner, S.A., 2003b, "Light Propagation in the Gravitational Field of Moving
Bodies by means of Lorentz Transformation. I. Mass monopoles moving with
constant velocities", A \& A 404, 783.

Klioner, S.A., Kopeikin, S.M., 1992, "Microarcsecond Astrometry in Space:
Relativistic Effects and Reduction of Observations", AJ 104, 897.

Klioner, S.A., Peip, M., 2003, "Numerical simulations of the light
propagation in the gravitational field of moving bodies", A \& A 410, 1063.

Klioner, S.A., Zschocke, S., 2010, "Numerical versus analytical accuracy of
the formulas for light propagation", Class. Quantum Grav. 27, 075015.

Klioner, S.A,. Soffel, M., 2014, to be published

Kopeikin, S.M., 1997, "Propagation of light in the stationary field of a
multipole gravitational lens", J. Math. Phys. 38, 2587.

Kopeikin, S.M., Korobkov, P., 2005, "General Relativistic Theory of Light
Propagation in the Field of Radiative Gravitational Multipoles",
arXiv:gr-qc/0510084v1.

Kopeikin, S.M., Korobkov, P., Polnarev, A., 2006, "Propagation of light in
the field of stationary and radiative gravitational multipoles", Class.
Quantum Grav. 23, 4299.

Kopeikin, S.M., Makarov, V.V., 2007, "Gravitational bending of light by
planetary multipoles and its measurement with microarcsecond astronomical
interferometers", Phys. Rev. D 75, 062002.

Kopeikin, S.M., Mashhoon, B., 2002, "Gravitomagnetic-effects in the
propagation of electromagnetic waves in variable gravitational fields of
arbitrary-moving and spinning bodies", Phys. Rev. D 65, 064025.

Kopeikin, S.M., Sch\"afer, G., 1999, "Lorentz covariant theory of light
propagation in gravitational fields of arbitrary-moving bodies" Phys. Rev. D
60, 124002.

Kopeikin, S.M., Sch\"afer, G., Gwinn, C.R., Eubanks, T.M., 1999, "Astrometric
and timing effects of gravitational waves from localized sources", Phys. Rev.
D 59, 084023.

Le Poncin-Lafitte, C., Linet, B., Teyssandier, P., 2004, "World function and
time transfer: general post-Minkowskian expansions", Class. Quantum Grav. 21,
4463.

Le Poncin-Lafitte, C., Teyssandier, P., 2008, "Influence of mass multipole
moments on the deflection of a light ray by an isolated axisymmetric body",
Phys. Rev. D 77, 044029.

Minazzoli, O.L., Chauvineau, B., 2009, "Post-Newtonian metric of general relativity including  
all the $c^{-4}$ terms in the continuity of the IAU2000 resolutions",  
Phys. Rev. D 79, 084027.  

Teyssandier, P., Le Poncin-Lafitte, C., 2008, "General post-Minkowskian
expansion of time transfer functions", Class. Quantum Grav. 25, 145020.

Thorne, K.S., 1980, "Multipole expansions of gravitational radiation", Rev.
Mod. Phys. 52, 299.

Xu, C., Wu, X., 2003, "Extending the First-Order Post-Newtonian Scheme in Multiple Systems to the 
Second-Order Contributions to Light Propagation", Chinese Physics Letters 20, No. 2, 195.  

Zschocke, S., Klioner, S.A., 2011, "On the efficient computation of the
quadrupole light deflection", Class. Quantum Grav. 28, 015009.

}

\end{document}